\documentclass[12pt]{iopart}
\newcommand{\be}{\begin{equation}}
\newcommand{\ee}{\end{equation}}
\newcommand{\bea}{\begin{eqnarray}}
\newcommand{\eea}{\end{eqnarray}}

\usepackage{epsfig}
\usepackage{amssymb}
\usepackage{units}
\usepackage{cleveref}
\usepackage{iopams}
\usepackage{setstack}
\usepackage{harvard}
\usepackage{color}

\begin{document}
\title{Polar molecule reactive collisions in quasi-1D systems}

\author{A. Simoni$^a$, S. Srinivasan$^a$, J-M. Launay$^a$, K. Jachymski$^b$, Z. Idziaszek$^b$, and~P.~S.~Julienne$^c$}
\address{$^a$Institut de Physique de Rennes, UMR 6251 du CNRS and Universit\'e de
Rennes 1, 35042 Rennes Cedex, France}
\address{
$^b$ Faculty of Physics, University of Warsaw, Pasteura 5,
02-093 Warsaw, Poland
}
\address{$^c$
Joint Quantum Institute, University of Maryland and National Institute of Standards and Technology, College Park, Maryland 20742, USA}


\begin{abstract}
We study polar molecule scattering in quasi-one-dimensional geometries.
Elastic and reactive collision rates are computed as a function of
collision energy and electric dipole moment for different confinement
strengths. The numerical results are interpreted in terms of first order
scattering and of adiabatic models. Universal dipolar scattering is
also discussed. Our results are relevant to experiments where
control of the collision dynamics through one dimensional confinement
and an applied electric field is envisioned.

\end{abstract}


\maketitle

\section*{Introduction}

Ultracold gases of polar molecules generate considerable theoretical
and experimental interest~\cite{Quemener2012}. Systems composed of polar molecules in
optical lattices are expected to give rise to a variety of novel quantum
phases~\cite{KG-2002-PRL-170406,SY-2007-PRL-260405,Sinha2007,BCS-2010-PRL-125301,Wall2010,Gorshkov2011,Sowinski2012}.
The presence of dipolar confinement-induced resonances has
been theoretically addressed for nonreactive molecules in reduced
quasi-1D geometries~\cite{2013-PG-PRL-183201,Bartolo}. Understanding reactive
processes is also relevant in quantum chemistry at extremely low
temperature~\cite{RK-2008-PCCP-4079,LDC-2009-NJP-055409}. For instance,
in three dimensions a fast chemical rate between ultracold fermionic
K-Rb molecules has been found~\cite{2010-SO-SCI-853}, which rapidly
increases even further if the molecules are polarized by an external
electric field~\cite{2010-KKN-NAT-1324}. These results have been
theoretically explained based on effective two-body quantum reaction
models~\cite{GQ-2010-PRA-022702,ZI-2010-PRL-113202}.

Unfortunately, at least for reactive species, reaction events are usually
responsible for a short lifetime of the trapped molecular sample which
may impair the observation of interesting collective phenomena. In
order to overcome this limitation, scattering of reactive molecules in
two-dimensional pancake shaped traps has been experimentally studied
in~\cite{2011-MHGDM-NP-502}. In this geometry, the repulsion between
molecules with dipoles parallel to the trap axis allows the reactive
rate constants to be suppressed, as confirmed by quantum theoretical
calculations~\cite{GQ-2010-PRA-060701,AM-2010-PRL-073202}.

Recent experimental work addressing fermionic K-Rb molecules in quasi-1D
systems shows that reactive processes are very sensitive to both the
intensity of polarizing electric fields and to the depth of an additional
lattice along the system axis~\cite{2012-AC-PRL-080405}. However, the
amount of data concerning the influence of the electric field on the
scattering dynamics reported in~\cite{2012-AC-PRL-080405} is relatively
limited. Motivated by ongoing experimental studies on ultracold K-Rb
and other rapidly developing experiments with ultracold molecules, in
the present work we address theoretically scattering of reactive polar
molecules in quasi 1D geometry achieved by tight lattice confinement in
two directions at very low collision energy for a variety of trapping
and molecular interaction parameters. We focus on the cases of identical fermions or
bosons, which are both achievable experimentally. We discuss emergence of universal scattering
in 1D configurations for strong dipole moments. As we restrict our considerations 
to a two-body scattering problem, we will not take into account the many-body 
correlations which can have strong impact on the dissipative dynamics of quasi-one-dimensional 
gases~\cite{Syassen2008,Zhu2014}.

The paper is organized as follows. We first present the theoretical
model and our computational approach. We then summarize and discuss our
results for different confinement strengths as a function of collision
energy, dipole moment and orientation. When possible we interpret the
results based on analytical or effective models.  A short conclusion
ends this work.


\section{Model}
\label{model}

The short range dynamics of a molecule-molecule collision is a complex
reactive four-body problem very difficult to treat theoretically, in
particular in the ultracold regime. In this work we follow the approach
of~\cite{AM-2010-PRL-073202} by limiting ourselves to an asymptotic model
where the collision process is described as an effective two-body problem
at intermolecular distances larger than $a_c$ which denotes the distance
where atom exchange processes take place. Then the reaction phenomena can simply be
taken into account by introducing a WKB-like boundary condition
that guarantees a given fraction of the flux reaching the short range $r \sim a_c$
to give rise to reaction~\cite{AM-2010-PRL-073202}. Such fraction is an
empirical parameter in the model and will be taken to be unity (total absorption) in the
present work as suggested by experiments performed on fermionic K-Rb
molecules~\cite{2010-SO-SCI-853}. This approach corresponds to the so-called universal regime where short range interaction details do not influence the dynamics. The model can easily be generalized to nonuniversal cases as well~\cite{ZI-2010-PRL-113202,Jachymski2013}.

The presence of external fields induce an effective dipole moment
in the molecules, which we will treat as fixed along the field axis.
After separation of the center of mass, the potential for two fixed
dipoles ${\vec d}= d {\hat n}$ oriented along the $\hat n$ direction
and subject to 2D harmonic confinement with frequency $ \nu_\perp$ reads
\be
\label{ham}
 H=   -\frac{\hbar^2}{2 \mu} {\Delta} + \frac{1}{2} \mu \omega_\perp^2 \rho^2 + V_{int}({\vec r}) ,
\ee
with $\rho$ the distance to the trap axis $z$ and $\omega_\perp=2 \pi \nu_\perp$ the angular oscillation frequency.
In the outer region of radial distances $r>a_c$ the intermolecular
interaction $V_{int}$ takes a simple form comprising to leading order
the isotropic van der Waals interaction and the anisotropic potential
between oriented dipoles ${\vec d}= d {\hat n}$
\be
 V_{ int}({\vec r}) = V_{vdW} + V_{\rm dd} = -\frac{C_6}{r^6} + \frac{d^2}{r^3} \left[ 1 -3
 ({\hat n} \cdot {\hat r})^2   \right] .
\ee
We ignore in this work the correlated fluctuations of molecular dipoles at
short range which can become important and give rise to resonant effects
as the molecular dipole increases~\cite{PSJ-2011-PCCP-19114}. Within the present
model, the importance of the different terms in the Hamiltonian can be
summarized in typical natural lenghscales. The average scattering length
for a van der Waals potential ${\bar a}= 2 \pi / \Gamma^2(1/4)  (2 \mu
C_6 / \hbar^2)^{1/4} $ first introduced in~\cite{GFG-1993-PRA-546}, the dipolar
length $a_{\rm dd}=\mu d^2/ \hbar^2$, and the harmonic oscillator length
$a_{\rm ho}=\sqrt{\hbar / \mu \omega_\perp }$ will be used respectively
to estimate the range of the isotropic, anisotropic dipolar, and of the
harmonic confinement potentials. Sample values of the trapping and molecular parameters
are reported for reference in table~\ref{table1} for reactive molecules with a mild (K-Rb)
and strong (Li-Cs) proper dipole moment.
\begin{table}[t]
\begin{center}
\caption{
Characteristic trapping and potential parameters as defined in
the text for two sample reactive polar molecules with relatively small
(K-Rb) and large (Li-Cs) electric dipole moment. Three sets of
parameters corresponding to the weak, intermediate and strong confinement
cases discussed in the text are provided. Two specific values of electric dipole $d_1$ and $d_2$ at
which $a_{\rm dd}={\bar a}$ and $a_{\rm dd}=a_{ho}$ are also given for reference. 
Van der Waals coefficients are from~\cite{SK-2010-NJP-073041} and~\cite{PSZ-2013-PRA-022706} for K-Rb and Li-Cs, respectively.
Note that for the sake of completeness values larger than the permanent K-Rb dipole moment  
of 0.57~D have also been used in the calculations.
}
\label{table1} \vskip
12pt
\begin{tabular}{|l |  l l l l l l|}
\hline \hline
{\rm System}  &    ${\bar a} (a_0)$ &  $a_{ho} (a_0)$  & $\nu_\perp$(MHz)  &  $\hbar \omega_\perp/ k_B (\mu{\rm K})$ & $d_1$(D)& $d_2$(D) \\
\hline
{\rm K-Rb} & $118$  & $1.18\times 10^4$    &  $4.08\times 10^{-4}$   & $1.96\times 10^{-2}$ & $8.12\times 10^{-2}$  &  $8.12\times 10^{-1}$  \\
           &            & $1.18\times 10^3$    &  $4.08\times 10^{-2}$   & 1.96                 &           &  $2.57\times 10^{-1}$  \\
           &            & $1.18\times 10^2$    &  $4.08 $       & $1.96\times 10^{2}$           &           &  $8.12\times 10^{-2}$  \\
{\rm Li-Cs} & $256$  & $2.56\times 10^4$   &  $8.24\times 10^{-4}$    & $3.95\times 10^{-2}$ & $3.69\times 10^{-1}$ &  3.69 \\
            &            & $2.56\times 10^3$   &  $8.24\times 10^{-2}$    & 3.95     & & 1.17  \\
            &            & $2.56\times 10^2$   &  $8.24$       & $3.95\times 10^{2}$  & & $3.69\times 10^{-1}$ \\

 \hline \hline
\end{tabular}
\end{center}
\end{table}

The collision process will be described in terms of the scattering matrix elements
$S_{j^\prime j}$ between the eigenstates $ |j \rangle = | n_r \Lambda
\rangle $ of the transverse quantum oscillator in the $\rho$ direction
with radial quantum number $n_r=0,1,\dots$, axial angular momentum $\Lambda$,
and energy $E_{n_r \Lambda}= \hbar \omega_{\perp} (2 n_r + |\Lambda | +1 )$.
The asymptotic scattering wavefunction for bosonic and fermionic exchange symmetry
can be expressed in terms of ingoing and outgoing waves along the trap axis as
\be
\Psi^{(j) {\rm B,F}}(r) \sim {q_j^{-1/2}}  e^{-i q_j |z|} | j \rangle
\pm \sum_{j^\prime} q_{j^\prime}^{-1/2}  e^{i q_{j^\prime} |z|} S^{\rm B,F}_{j^\prime j} | j^\prime \rangle
\quad,\quad r \to \infty
  \label{psiasym}
\ee
where $q_j=\sqrt{2 \mu (E-E_j)}/\hbar$ is the wavevector in channel $j$.
The upper (lower) sign holding for bosons (fermions) in~(\ref{psiasym})
guarantees that in the absence of interaction $S_{j^\prime j}
=\delta_{j^\prime j}$. Finally, the inelastic state-to-state collision rate ${\cal K}_{j^\prime
j}$ for  $ |j \rangle  \to   |j^\prime \rangle$ transitions is expressed in
terms of the $S$ matrix elements as ${\cal K}^{\rm B,F}_{j^\prime
j} = v_j  \left| \delta_{j^\prime j} - S_{j^\prime j}^{\rm B,F}  \right|^2 $,
with $v_j=\hbar q_j / \mu$ the relative velocity in channel $j$.

\subsection{Wigner laws and Born approximation}

The effective potential between polar molecules has a long-range
inverse-cube character that profoundly modifies the usual Wigner
laws valid for short range potentials~\cite{Sadeghpour2000}. The corresponding asymptotic
low energy expansion of the elastic phase shift has been derived
in~\cite{BG-1999-PRA-2778,TOM-2013-PRL-260401} for isotropic
repulsive and attractive inverse cube potentials, respectively.
In order to generalize these results to the case of a complex boundary
condition, we first define a complex phase shift for bosons and fermions
as $\tan(\delta^{\rm B,F}) = i (1-S_{jj}^{\rm B,F}) / (1+S_{jj}^{\rm
B,F})$~\cite{AM-2010-PRL-073202}, where the channel index $j$ on the lhs
has been, and will henceforth be unless needed, left implicit for notational convenience.
We can then express the phase shift as
\be
\label{tan1}
\tan(\delta^{\rm F}) =   -(q b_3) \log(|b_3| q ) - q (\alpha^{\rm F} -i \beta^{\rm F})
\ee
and
\be
\label{tan2}
\tan(\delta^{\rm B}) =  \frac{1}{ (q b_3) \log(|b_3| q ) + q (\alpha^{\rm B} -i \beta^{\rm B}) }
\ee
in terms of an algebraic length $b_3=  2 a_{\rm dd} \left [3 ({\hat n}
\cdot {\hat z})^2 -1\right] $ that characterizes the strength of the dipolar
interaction for oriented molecules.
This definition corresponds to an asymptotically attractive (repulsive)
dipolar interaction $-b_3/z^3$ for $b_3>0$ ($b_3<0$).

The leading order $q \log(|b_3| q)$ is universal in that it only depends
on the long-range dipolar interaction. Its value can also be obtained
in the Born approximation which can be shown to be exact in the zero
energy limit for long range potentials~\cite{1960-LMD-NP-275}. The next
order energy correction is proportional to the complex length $(\alpha
-i \beta)$ which, in addition to $\vec d$, also depends on the details
of the short-range dynamics, with the imaginary part $\beta$ accounting
in particular for inelastic and reactive processes. We will often term
in the following $\alpha$ and $\beta$ the elastic and inelastic
collision lengths, respectively.

As the dipolar force becomes dominating the purely dipolar
model~\cite{TOM-2013-PRL-260401} shows that for $b_3 >0$ the scaling
of $\alpha$ and $\beta$ is set by $a_{\rm dd}$. Similarly, in the $b_3< 0$
regime according to~\cite{BG-1999-PRA-2778} one may expect $\alpha \sim
a_{\rm dd}$. However, we anticipate that since for $b_3< 0$ the dipolar
repulsive potential creates a barrier to the reaction, depending on
collision energy and on the confinement strength the inelastic length
$\beta$ may not scale with $a_{\rm dd}$ and can be exponentially suppressed.

If $(\alpha -i \beta)$ is considered to be a momentum-dependent quantity
(finite for $q\to 0$) the parametrizations (\ref{tan1}) and (\ref{tan2})
are indeed exact. The corresponding elastic scattering rates can be expressed using the
general formulas in~\cite{PN-2006-PRA-062713} as
\be
\label{kelb}
{\cal K}^{\rm el,B} \equiv  {\cal K}_{jj}^{\rm B} =
\frac{4 \hbar q}{\mu} \frac{1}{q^2 \left[\alpha^{\rm B} + b_3 \log(|b_3| q )   \right]^2 + (1+q \beta^{\rm B})^2}
\ee
and
\be
\label{kelf}
{\cal K}^{\rm el,F} \equiv  {\cal K}_{jj}^{\rm F}  =
\frac{4 \hbar q^3}{\mu} \frac{\left[ \alpha^{\rm F} + b_3 \log(|b_3| q )   \right]^2
+ \left( \beta^{\rm F}\right) ^2}{q^2 \left[\alpha^{\rm F} + b_3 \log(|b_3| q )   \right]^2 + (1+q \beta^{\rm F})^2}
\ee
for bosons and fermions, respectively.
The total inelastic plus reactive rate from channel $j$ takes the form
\be
\label{kinel}
{\cal K}^{\rm inel} = v_{j}  ( 1 - |S_{jj}^{\rm B,F}|^2  )  = \frac{4 \hbar q^2}{\mu}
\frac{\beta^{\rm B,F}}{q^2 \left[\alpha^{\rm B,F} + b_3 \log(|b_3| q )   \right]^2 + (1+q \beta^{\rm B,F})^2}
\ee
for bosons and fermions alike.
The purely reactive rate for molecules in channel $j$ is finally expressed
through the lack of unitarity of the $S$ matrix
as $ {\cal K}^{\rm react} =  v_j ( 1-\sum_{j^\prime}   |S_{j^\prime j} |^2 ) $.
In striking contrast with the 3D case, for bosons in one dimension the elastic
scattering rate will dominate the reactive rate if the collision energy is
sufficiently small.
The effect of the full potential at collision energy
$E$ can also be conveniently summarized in a complex one dimensional
scattering length~\cite{1998-MO-PRL-938,2002-ELB-PRA-013403}, defined as
\be
a_{1D}^{\rm B}(q)=\lim_{q \to 0} \frac{1}{q \tan(\delta^{\rm B})} \quad , \quad
a_{1D}^{\rm F}(q)=-\lim_{q \to 0} \frac{\tan(\delta^{\rm F})}{q} .
\ee
Using \cref{tan1,tan2} one promptly obtains
\be
\label{a1D}
a_{1D}^{\rm B,F}(q)= b_3 \log(|b_3| q ) + \alpha^{\rm B,F} -i \beta^{\rm B,F} .
\ee
In order to clarify the rationale for the different conventions adopted
so far, we remark that for purely elastic scattering in a short-range
potential the zero-energy wave function will take at long range the
familiar form $\Psi \sim (|z| - a_{1D})$ for both bosons and fermions.
Note however that if $d  \neq 0$, the logarithmic singularity in
\cref{a1D} makes $a_{1D}$ an intrinsically energy-dependent quantity
that does not admit a finite zero-energy limit. It will be shown in the
following that such Wigner laws set in at energies well below the typical
current experimental conditions and may not be directly observable.

In 3D, elastic collisions in the presence of dipole-dipole interactions are well described using Born approximation. Similar approach has been already applied in quasi-2D geometry as well~\cite{AM-2010-PRL-073202}. In the present quasi-one-dimensional case, we will make use of the distorted wave Born approximation (DWBA) for interpretation purposes.

In this method we first calculate the scattering amplitude and the wave function for van der Waals interaction alone using standard Born approximation and then include the dipole-dipole interactions on top of it.  The van der Waals scattering amplitudes within our definitions yield
\be
\label{amplt_bf}
f_{\rm vdW}^{\rm B}(q)=-\frac{1}{i q a^{\rm B}_{1D}(q)+1}\quad,\quad f_{\rm vdW}^{\rm F}(q)=\frac{1}{i/(q a^{\rm F}_{1D}(q))-1},
\ee
where in the limit of $\bar{a}\ll a_{ho}$ and $q\to 0$ we have~\cite{AM-2010-PRL-073202}
\be
\label{a1Dan}
a^{\rm B}_{1D}(q)\stackrel{q\to 0}{\rightarrow}-\frac{a_{\rm ho}^2}{4\bar{a}}(1+i) \quad , \quad
a^{\rm F}_{1D}(q)\stackrel{q\to 0}{\rightarrow}-\frac{6 \bar{a}^2\bar{a}_{1}}{a_{\rm ho}^2}(1+i).
\ee
Here ${\bar a}_1=\Gamma(1/4)^6/(144\pi^2\Gamma(3/4)^2){\bar a}\approx 1.064{\bar a}$~\cite{ZI-2010-PRL-113202}. The total scattering amplitude in DWBA is given as a sum $f=f_{\rm vdW}+f_{\rm dd}$, where $f_{\rm dd}$ is now calculated including the phase shift from van der Waals interaction
\be
f_{\rm dd}^{\rm B,F}(q) = -4 i e^{2i\delta_{\rm vdW}^{\rm B,F}(q)}\frac{a_{\rm dd}}{q a_{\rm ho}}I^{\rm B,F}(q),
\ee
where
\be
\label{IF}
I^{\rm B}(q) = \int_0^{\infty} d \rho \int_{0}^{\infty} dz \cos^2 (q z+\delta_{\rm vdW}^{\rm B}) e^{-\rho^2}
\frac{z^2-\rho^2/2}{(z^2+\rho^2)^{5/2}}
\ee
and for fermions the $\cos$ under the integral has to be replaced with $\sin$. This formula assumes that the particles are in the ground state of transverse harmonic oscillator and can easily be generalized by replacing the ground state wave function with an excited one. 
Finally equations~(\ref{amplt_bf}) are inverted to obtain from the full scattering amplitude $f$ the energy dependent scattering lengths.
\subsection{Numerical calculations}

Details on the computational algorithm used for the present problem
will be given elsewhere and we limit ourselves here to providing the
main elements.

In our approach the scattering equations are written in spherical
coordinates and the interaction volume is partitioned in two regions
in which the wavefunction is expanded on different basis sets. In the
internal region a basis of spherical harmonics $Y_{\ell m}$ is used.
In practical calculations we find that numerical convergence is optimal
if the small $r$ region is made to extend up to about 10~$a_{ho}$.  Note
that the Hamiltonian (\ref{ham}) is symmetric with respect to reflection
about each plane containing the trap axis and about the plane orthogonal
to the trap axis and containing the origin. The Hamiltonian matrix can
therefore be partitioned into eight noninteracting blocks labeled by
$\{ \pi_x \pi_y \pi_z \}$, with $\pi_{x,y,z} = \pm 1$ for positive and negative
parity, thus leading to significant computational saving. Moreover,
according to the tensor order 2 of the intermolecular interaction and
trapping potential, only states with $\Delta \ell =0,\pm2$ and  $\Delta
m =0,\pm2$ are coupled, resulting in a highly sparse interaction matrix.
Identical particle symmetry in the present spinless case is equivalent
to total parity $\pi=\pi_x \pi_y \pi_z$ and requires that even (odd) partial waves
only are allowed for identical bosons (fermions).

In the external region the cylindrical trapping potential tends to
confine the wavefunction to small angular regions near the poles, making
the development in spherical harmonics inefficient. In this region
we use therefore an adapted primitive basis obtained by projecting
cylindrical grid (DVR) functions on the spherical surface. Such basis
functions inherently depend on the distance $\rho$ to the trap axis and
therefore naturally adapt with varying $r$ to the cylindrical symmetry
of the problem.  Our grid excludes the pole $\theta=0$ in order to avoid
the singularity of the orbital angular momentum operator ${\vec \ell}^2$
at this point. The boundary condition imposed at the other grid edge is
that the wavefunction vanishes at (and beyond) a given distance to the
axis $\rho_{\rm max}$. Such DVR grid containing only one edge point is
known as of Gauss-Lobatto-Radau type.

The radial wavefunction is computed with exponential numerical accuracy
up to a distance $r_\infty$ using the spectral element approach we
recently adapted to molecular physics. In such approach the radial
domain is partitioned in bins and the scattering problem is cast as a
linear system for a sparse matrix which is solved using state-of-the-art
computer packages. One main advantage of our method is that the extremely
sparse nature of the interaction potential is fully taken advantage of.

Note that in the external region we do not work directly in the primitive
basis but find it convenient to use a diabatic-by-sector approach in
which the Hamiltonian is diagonalized at the center of each bin and the
wavefunction is developed (in the given bin) on a small number of the
resulting eigenvectors~\cite{1986-BL-CP-103}.

Finally, the reactance and hence the scattering matrices are extracted by
matching the wavefunction at $r_\infty$ to asymptotic reference functions
expressed in cylindrical coordinates. The explicit formulas and procedure
are a direct generalization of the ones in~\cite{2011-AS-JPB-235201}
where one must now also sum over different values of $\Lambda$. Moreover,
in order to enforce the correct Wigner law for an effective $z^{-3}$
potential it is essential to correct the trigonometric reference functions
using first order Born approximation. To this aim a semianalytical
procedure based on the evaluation of exponential integrals in the same
spirit of~\cite{GHR-1999-JCP-10418} has been implemented in our algorithm.


\section{Results and discussion}
\label{results}

In spite of the simplifying assumptions at the basis of our model, the
problem is still relatively complex as it depends on the three independent
length scales $\bar a$, $a_{ho}$, and $a_{\rm dd}$ arising from the different
interaction terms in the Hamiltonian, on the collision energy $E$, on the
molecular dipole orientation, and on the initial quantum state for the
collision. We will present our results according to a classification based
on the confinement strength of the trap, that will be referred to as weak,
intermediate, and strong, corresponding respectively to $a_{ho}=10^n
~{\bar a}$ with $n=2,1,0$. 
\begin{figure}[hb!]
\centering
\includegraphics[width=0.9\columnwidth]{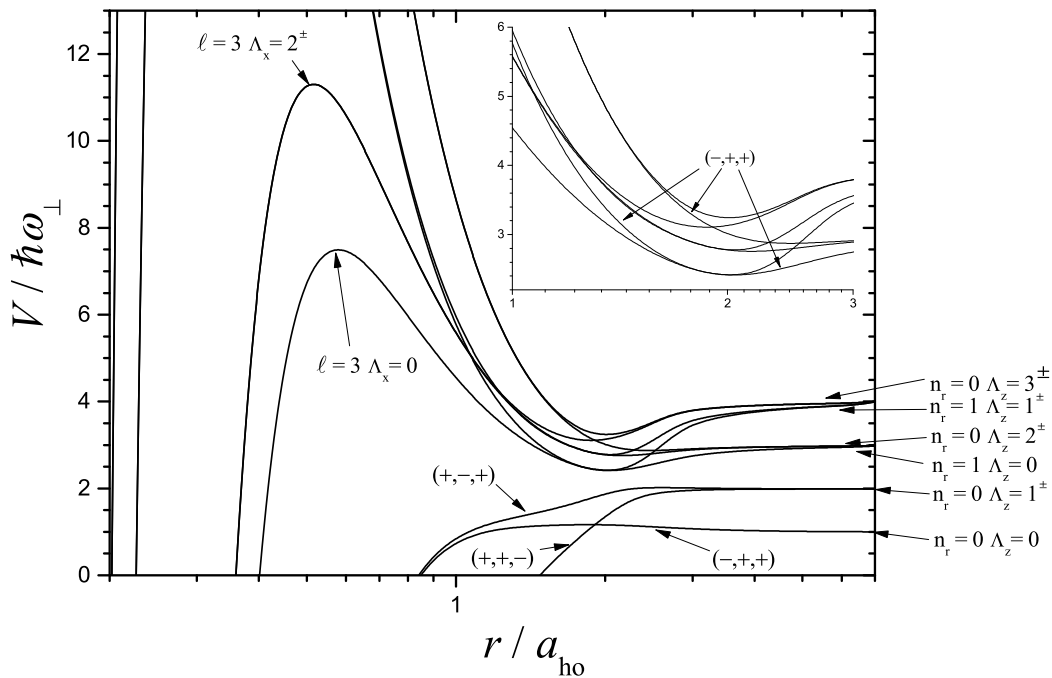}
\caption{\label{adiab} Adiabatic potential curves in spherical
coordinates for fermionic K-Rb molecules with $d=0.5~$D and
$\nu_\perp=50~$kHz. The $n_r$ denotes the radial quantum
number of 2D harmonic oscillator, $\Lambda_z$ the projection of the
angular momentum on the trap axis and $\Lambda_x$ on the dipoles axis. We
note that $\Lambda_z$ becomes a good quantum number at large distances,
and $\Lambda_x$ is approximately good only at distances where the trap can be neglected. The $\pm$ signs refer to the
reflection symmetry with respect to the sign change of each of the $(x,y,z)$ Cartesian coordinates.}
\end{figure}
Whereas this convention denotes
the strength of the confinement with respect to the isotropic part of
the potential, one should note from table~\ref{table1} that for ordinary
molecular parameters the dipolar length $a_{\rm dd}$ may easily become the
dominant length scale in the problem. 
We will mainly focus on the case when the orientation of the dipoles is perpendicular to the trap axis, unless stated otherwise. 
For the sake of a more direct
comparison with experiments most of our results will be expressed in
dimensional form using K-Rb physical parameters. For completeness, we
will however also consider values of $d$ larger than the proper physical
dipole moment of K-Rb. Since the model is universal, if needed it is
easy to transpose the present data to other polar molecule systems by
dimensional scaling.

To interpret our numerical results, we will make use of simpler models. DWBA will be applied to describe the elastic rates. In order to understand the behavior of inelastic rates, one can use the adiabatic approximation. In this method one aims to partially diagonalise the problem. Depending on the chosen coordinate system, one either diagonalises the angular part of the wave function at each distance $r$ or the $(\rho,\phi)$ dependent part at each distance $z$. The former choice better reflects the symmetry of the interaction, while the latter is more suited to the trapping potential, but neither will fully cover the all the features of the system. Below we use spherical coordinates. 
The partial diagonalisation results in a set of adiabatic curves $\lambda_n(r)$, which act as an effective potential that can be plugged into radial Schr\"{o}dinger equation.
This method cannot be expected to give strictly quantitative results for the reaction rates, but nicely shows the couplings between different states. Figure~\ref{adiab} shows an exemplary set of adiabatic curves for identical fermions with $d=0.5~$D confined in a trap with frequency $\nu_\perp=50~$kHz, giving $a_{ho}=3.6~{\bar a}$, between the medium and strong confinement case. At large distances, when the dipole-dipole interaction becomes negligible, the adiabatic potentials correspond to the states of two-dimensional harmonic oscillator. At smaller $r$ the interplay between the centrifugal barrier and the interaction becomes crucial, which is why spherical coordinates seem to be the better choice here. One can see that strong dipolar interactions can completely remove the barrier for the lowest adiabatic curve.
\begin{figure}[ht!]
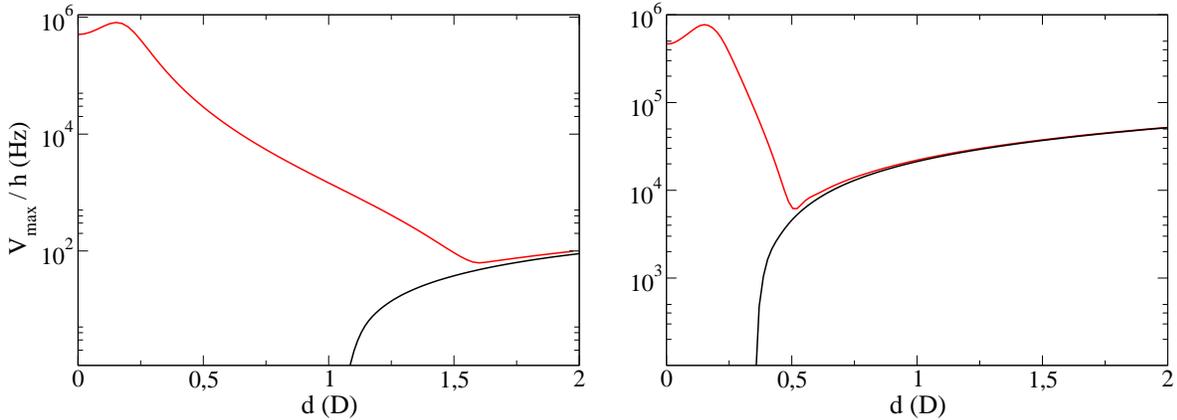

\centering
\includegraphics[angle=0,width=0.49\columnwidth] {fig2a}
\includegraphics[angle=0,width=0.49\columnwidth] {fig2b}
\caption{\label{abh}
Adiabatic barrier heights for weak (left) and intermediate (right) confinement corresponding to $a_{\rm ho}=100{\bar a}$ and $10{\bar a}$, respectively, for bosons (black curves) and fermions (red curves). The reduced mass and $C_6$ of K-Rb molecules is assumed.}
\end{figure}
To get an estimate of the rate constant in the energetically lowest channel we then restrict ourselves to the lowest adiabatic curve, imposing MQDT boundary conditions at short range. This neglects the nonadiabatic couplings, but still qualitatively reproduces the full numerical results, as will be shown in the next sections. This is the case because the main factor determining the rates is the height of the adiabatic barrier that has to be overcome by the particles. Adiabatic barrier heights for both bosons and fermions at low and medium confinement are given in figure~\ref{abh}. Obviously, at zero dipole moment the barrier is present only for fermions. For strong dipoles the barrier height becomes independent of the quantum statistics, as the dipole-dipole interaction starts to dominate over the centrifugal barrier.

\subsection{Weak confinement}

Let us now consider scattering in a weakly confining trap $a_{ho}=10^2
~{\bar a}$ for molecules in the lowest transverse trap level
$| n_r \Lambda \rangle = | 0 0 \rangle$. For vanishing
$d$ and $E\to 0$ the scattering problem can be solved essentially exactly relying on the
separation of length scales $a_c \ll {\bar a} \ll a_{ho}$~\cite{AM-2010-PRL-073202}. Under
these conditions motion at short range is unaffected by the trap and is
driven by the spherically symmetric van der Waals interaction. Moreover,
the low collision energy in the axial direction and the small zero point
energy $\hbar \omega_\perp$ guarantee that total energy in the short-range
three dimensional motion will be small, such that interactions will be $s$-wave dominated.
The resulting 1D scattering lengths are given in equations~(\ref{a1Dan}). The very good agreement we obtain between the numerical and the analytical
results for both bosons and fermions confirms the physical picture
at the basis of (\ref{a1Dan}) and represents a stringent test for our
computational approach \footnote{For a fully quantitative comparison small
corrections to~(\ref{a1Dan}) of the order ${\bar a}/a_{ho}$ have been
taken into account; see equation (3) of~\cite{AM-2010-PRL-073202}.}. Note
that the condition ${\bar a} \ll a_{\rm ho}$ implies $\alpha^{\rm
F}/\alpha^{\rm B}=\beta^{\rm F} / \beta^{\rm B} \sim ({\bar a} / {a_{\rm
ho}})^4  \ll 1$ in the present case of weak confinement.
\begin{figure}[bt]
\includegraphics[width=0.9\columnwidth] {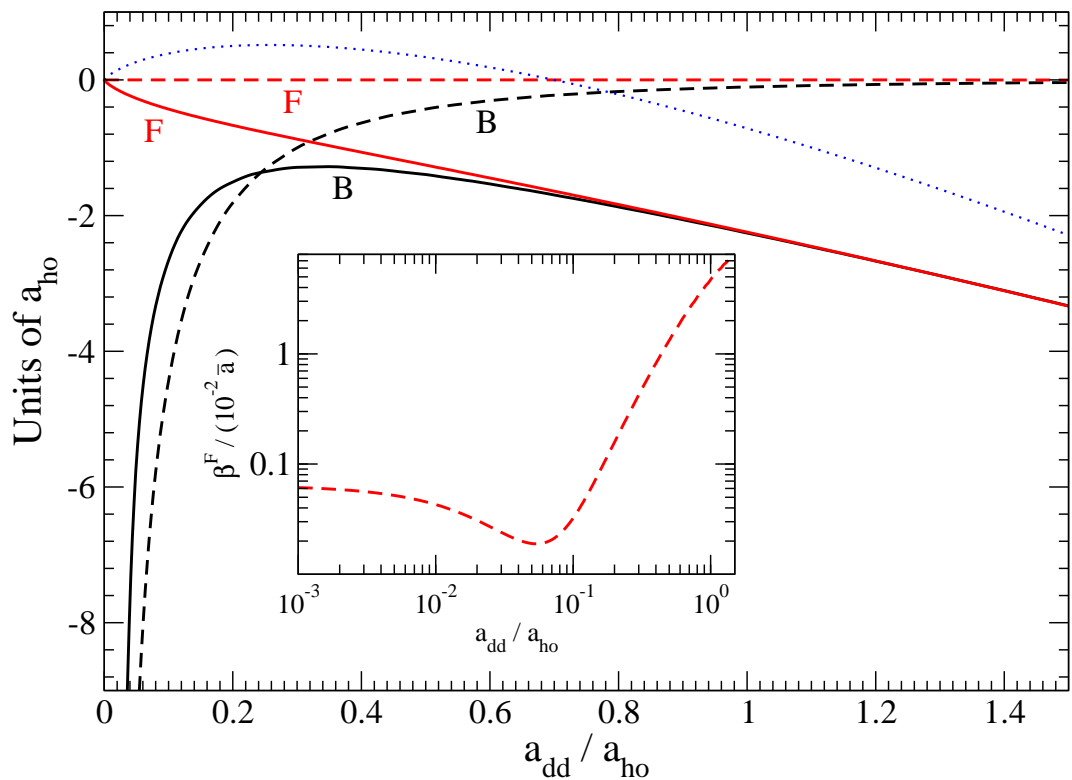}
\caption{
Main panel : the collision lengths $\alpha$ (full lines) and $
-\beta$ (dashed lines) defined in equations~(\ref{tan1}) and (\ref{tan2}), calculated numerically in the $E\to 0$
limit for bosonic (B) and fermionic (F) polar molecules as a function of
the dipolar length $a_{\rm dd}$. Transverse confinement $a_{ho}=100~ {\bar
a}$ is weak. The universal dipolar contribution $b_3 \log(q |b_3|)$
is also shown (dotted line) for a collision energy of $E/k_B=5$~nK and the reduced mass of K-Rb.
Inset : enlarged view of $\beta^{\rm F}$ in log-log scale. Note
the different units used for the vertical axis in the two panels.
}
\label{fig1}
\end{figure}

Let us now consider the behavior of $a_{1D}$ for nonvanishing $d$. In
the aim of suppressing harmful reactive collisions the molecular dipoles
are supposed to be orthogonal to the trap axis and calculations are
performed for reference in the zero energy limit. We extract the collision
lengths $\alpha$ and $\beta$ from the numerically calculated $a_{1D}$
using~(\ref{a1D}). As already remarked below equation~(\ref{tan2}),
the scaling behavior of $\alpha^{\rm B,F}$ can be expected to evolve
from the combination of $\bar a$ and $a_{\rm ho}$ given by~(\ref{a1Dan})
for small dipoles to $a_{\rm dd}$ in the $a_{\rm dd} \gg a_{\rm ho}$ limit. This
scaling argument is consistent with the extremely rapid increase of
$\alpha^{\rm F}$ with $d$ from small values $\sim 10^{-3} ~{\bar a}$ to
an approximate scaling with $a_{\rm dd}$ that sets in already for $a_{\rm dd}
\approx a_{\rm ho}$. On the converse, the bosonic analogue $\alpha^{\rm B}$
is initially large ($\alpha^{\rm B} = 25~ a_{\rm ho}$) and {\it
drops} to values on the order of $a_{\rm dd}$ for $a_{\rm dd} \approx a_{\rm ho}$.

\begin{figure}[ht]
\includegraphics[width=0.9\columnwidth] {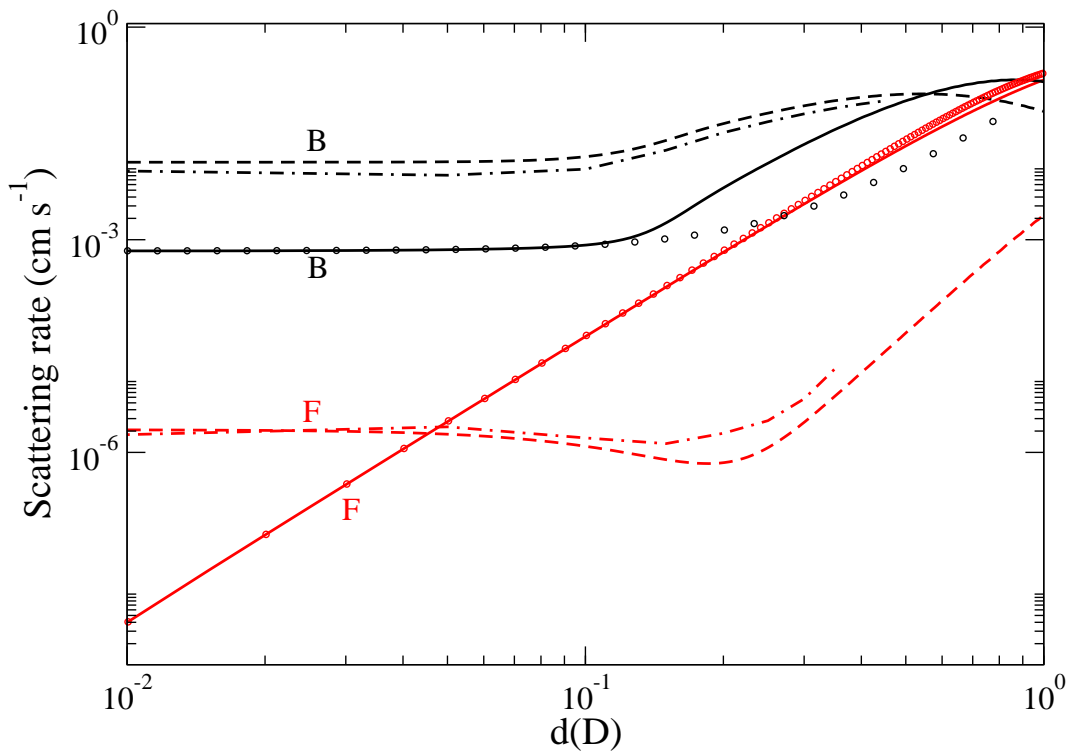}
\caption{
Numerically calculated elastic and inelastic collision rates for polar
molecules in quasi-1D geometries as a function of the induced dipole
moment at a collision energy $E/k_B=5~$nK. Transverse confinement $a_{\rm ho}=100~ {\bar a}$ is weak and K-Rb molecular parameters have been used.
Elastic (full lines) as well as reactive
rates (dashed lines) are shown for bosons (B) and fermions (F).
The points indicate the result of
distorted wave Born approximation to the elastic collision rates (see text). The dot-dashed lines represent the reactive 
rates calculated for the lowest adiabatic potential determined by 
diagonalization of the full matrix potential in the spherical basis (see 
text for details)
}
\label{fig2}
\end{figure}

Note from~(\ref{kinel}) that in the $E \to 0$ limit one simply
has $\cal K^{\rm inel} \propto \beta$. The behavior of $\beta$ with $d$
represented in figure \ref{fig1} can thus be interpreted in terms of
the reactive rate, {\it i.e.} of the flux that reaches the spherical
surface of radius $a_c$ and that in our model is fully adsorbed. From this
viewpoint, for collisions of bosonic molecules that do not present any
barrier at short range the long-range repulsion occurring at perpendicular
configuration for $d>0$ will directly tend to suppress the reaction. More
quantitatively, the minimum energy path connects a repulsive and an
attractive potential valley crossing a saddle point $\vec r^*$ where the
net force vanishes. It can be easily found that this point is located
in the plane of the dipoles at a distance $r^* =  6^{1/5} a_{\rm ho}^{4/5}
a_{\rm dd}^{1/5}$ from the origin with polar angle $\sin^2(\theta^*)=1/5$. The
potential energy barrier at the saddle point is $d^2 / {r^*}^2$. Tunneling
through the barrier results in the strong decrease of $\beta^{\rm B}$
with $a_{\rm dd}$ observed in figure~\ref{fig1}.

The reaction dynamics of fermionic molecules is qualitatively different
as it is mainly controlled by the short-range potential barrier arising from
the combination of the van der Waals attraction and of the centrifugal
repulsion illustrated in figure~\ref{adiab}. It is the presence of this barrier that results in the
suppression of $\beta^{\rm F}$ both for vanishing and finite
$d$ with respect to the barrierless bosonic case; see the inset of
figure~\ref{fig1}. Note that the weak confinement condition ${\bar a}
\ll a_{\rm ho}$ implies that the harmonic potential has negligible influence
on the barrier shape and position. It can be easily shown that in the
absence of dipolar forces the energetically lowest effective barrier
with $\ell=1$ is located at $r_1= 2.3153 ~ {\bar a} $.

In a perturbative picture valid for weak dipoles, the $\ell=1$
barrier is modified by the dipolar interaction by an amount $\delta
E = \langle Y_{10} | V_{\rm dd}(r_1) | Y_{10} \rangle_\Omega = d^2 /
r_1^3$, where the average $ \langle \cdot \rangle_\Omega$ is taken on the
unit spherical surface. Since $\delta E > 0$, the dipolar force tends
to reinforce the barrier thus reducing the reactive rate, {\it i.e.}
$\beta$. In other terms, Fermi symmetrization prevents the molecules
from occupying the regions near the $z=0$ plane where the head-to-tail
attraction is stronger.  This perturbative picture begins to fail when
$\delta E \sim \hbar^2 /(2 \mu r_1^2)$ and the repulsion between the
fundamental and the excited effective potentials lowers the potential
barrier. In this situation the dipolar force becomes dominant over the
centrifugal interaction allowing the molecules to tunnel more efficiently
through the centrifugal barrier, leading thus to an increased $\beta
$. The combination of these large and small $d$ effects accounts
for the minimum in $\beta$ observed for $a_{\rm dd} \sim {\bar a}$
in figure~\ref{fig1} and is confirmed by the behavior of the adiabatic
barriers illustrated in figures~\ref{adiab} and~\ref{abh}.

Note in figure~\ref{fig1} that for $a_{\rm dd} \gtrsim a_{\rm ho}$ one has
$\beta^{\rm B,F} \ll \alpha^{\rm B,F} \sim a_{\rm dd}$ and to a good
approximation $\alpha^{\rm B} = \alpha^{\rm F}$. That is, for large
dipoles scattering is mainly elastic and the elastic scattering phases
of boson and fermions are very similar since they are accumulated at long
range where the effect of bosonic/fermionic statistics is immaterial. We
will come back in some more detail on such boson-fermion equivalence in
the following.

Let us now model more realistic experimental conditions in a finite
temperature gas. To this aim the collision energy will be fixed at
a small yet finite value $E/k_B=5$~nK, which for K-Rb molecules is
below the trap level spacing $2 \hbar \omega_\perp$; see table 1. At
these extreme yet experimentally attainable temperatures inelastic
transitions $ | 0 0 \rangle \to  | n_r^\prime \Lambda^\prime \rangle $
with $E_{n_r^\prime \Lambda^\prime} >E_{00}$ are energetically forbidden.
Figure \ref{fig2} shows that the elastic rate for bosons presents a
relatively weak dependence on $d$ with an overall variation of about
one order of magnitude for $d$ below $1$~D. Let us remark that
the general structure of~(\ref{kelb}) predicts no dependence
of $\cal K_{\rm el}$ on $d$ to the extent that $(\alpha q)^2,\alpha
b_3 q^2, (\beta q) \ll 1$. The variation of $\cal K_{\rm el}$  with $d$
and in particular the broad maximum observed for $ a_{\rm dd} \sim q^{-1}$
results therefore from energy-dependent terms in the denominator of
(\ref{kelb}), which are non-negligible even at our small collision energy.

The elastic rate in the case of fermionic molecules presents on the converse a dramatically
rapid increase. In fact,
generalizing the arguments invoked for $E \to 0$ to finite energy, the dipolar
length $a_{\rm dd}$ rapidly overcomes ${\bar a}$ in setting the scale of
$a_{1D}$ and thus of $\cal K$ through~(\ref{kelf}). 
On a more quantitative ground, DWBA
accurately describes the elastic collision rate for fermions in the investigated range of dipole moments, 
while for bosons higher order terms an play important role and the DWBA is not as accurate.


The reactive rate for bosons represented in figure \ref{fig2} can be
partly understood based on the zero-energy limit results, which show a
monotonically decreasing $\beta$. In fact, since we numerically obtain a
similar monotonic behavior also for the finite energy quantity $\beta(q)$
and at our collision energy $q \beta \sim 1$ the maximum is once again
due to finite-energy corrections in the denominator of (\ref{kinel}).
This kind of behavior was also predicted for quasi-2D dipolar collisions~\cite{PSJ-2011-PCCP-19114} and can intuitively be explained by the shape of the interaction potential.
Since repulsion affects more strongly
slower collisions the drop of $\cal K^{\rm inel}$ with $d$ and hence the
location of the maximum shifts at smaller $d$ for decreasing $E$. Note that in the present weak confinement geometry the reactive rate of bosonic molecules varies relatively little,
less than one order of magnitude for $d < 1$~D, and that the ratio of
elastic to inelastic collision is of order one and thus not favorable
for experiments where a long lifetime is needed.

\begin{figure}[hb!]
\includegraphics[width=0.9\columnwidth] {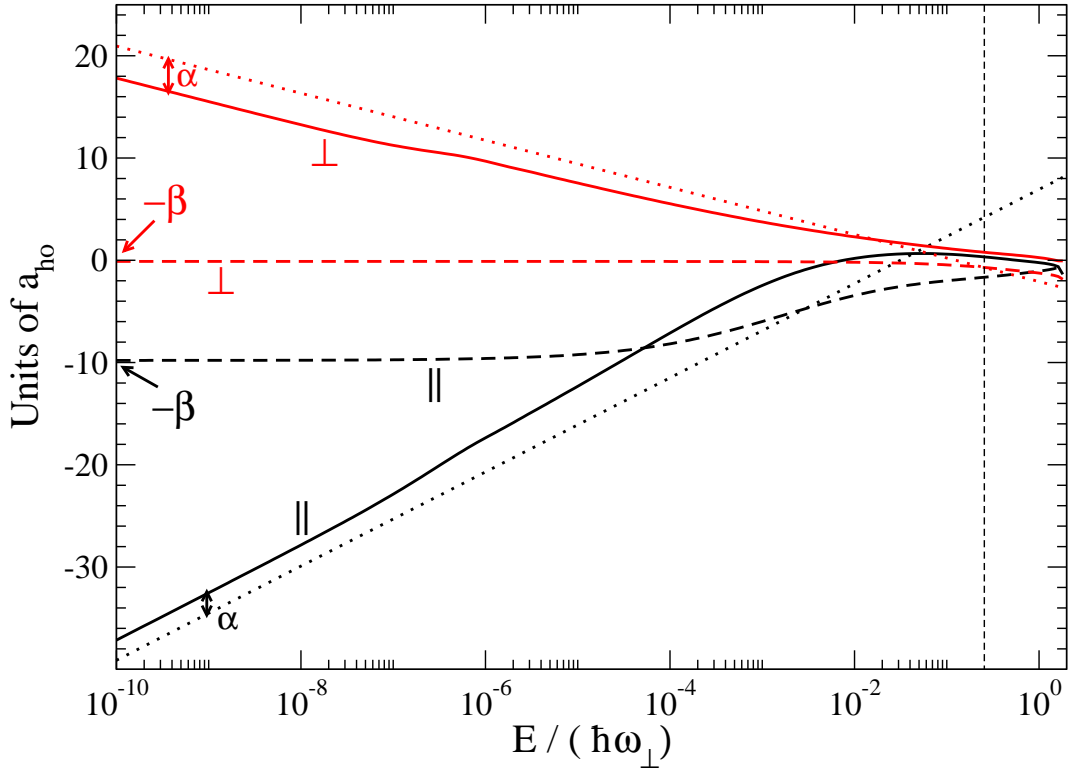}
\caption{
Real (full lines) and imaginary (dashed lines) parts of the 1D scattering
length as a function of collision energy for a dipole moment such that
$a_{\rm dd}=a_{\rm ho}$ in the case of intermediate confinement $a_{\rm ho}=10 ~
{\bar a}$.  Results are for bosonic molecules with dipole oriented
orthogonal $\perp$ or parallel $\|$ to the trap axis.  The dotted lines
represent the universal term $b_3 \log(b_3 |q|)$. The zero-energy limit
of the elastic and inelastic collision lengths $\alpha$ and $\beta$
is shown by arrows. The vertical line indicates a realistic experimental
temperature of $500~\mu$K.
}
\label{wignermd}
\end{figure}
To conclude, the inelastic rate for fermions in figure~\ref{fig2} presents
a modulation that closely mirrors the behavior of the zero-energy
inelastic length $\beta^{\rm F}$ in figure~\ref{fig1}. That is,
finite-energy corrections do not modify the reaction barrier arguments
used for $E \to 0$. Such arguments can then be repeated in the present case
to explain the presence of the shallow minimum for $d\approx 0.2$~D.
The ratio of inelastic to elastic collision rates becomes very favorable
for moderate dipole moments $d\sim 0.1$~D.  We have also checked that
as expected the fermionic reactive rate does not grow indefinitely with
$d$ since the long-range repulsion between dipoles perpendicular to the
trap will affect fermion dynamics on the same footing as it does for
bosons.


\subsection{Intermediate confinement}
We now consider a case with
stronger yet experimentally realizable confinement; $a_{\rm ho}=10 ~
{\bar a}$.  As expected, we find that the $d=0$, $E\to 0$ closed
expressions~(\ref{a1Dan}) valid for ${\bar a} \ll a_{\rm ho}$ do not
predict correctly the elastic collision length $\alpha$, which depends
on the detailed long-range interplay of trapping and van der Waals
potentials. 
\begin{figure}[!hbt]
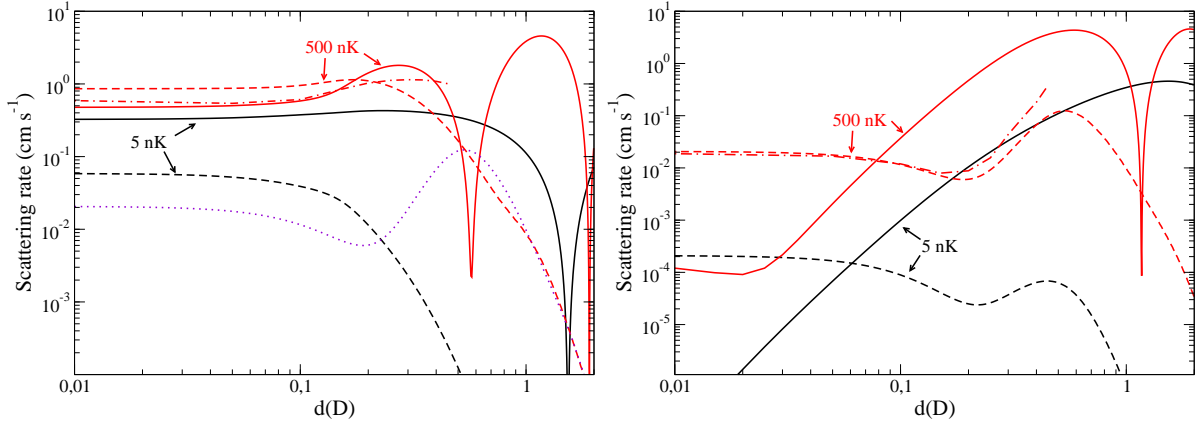

\centerline{
\epsfig{file=fig6a.eps,width=.5\columnwidth}
\epsfig{file=fig6b.eps,width=.5\columnwidth} }
\caption{Numerically
calculated elastic and inelastic collision rates for polar molecules
in quasi-1D geometries as a function of the induced dipole moment
for intermediate transverse confinement $a_{\rm ho}=10 {\bar a}$. Full (dashed)
lines in the leftmost panel
represent the elastic (inelastic) collision rates at energy $E/k_B=5$ and
$500$~nK for bosonic molecules. The rightmost panel shows the analogous quantities
for collisions of fermionic molecules. The dot-dashed lines on both panels show the results of the adiabatic approximation for $E/k_B=500~$nK. The fermionic inelastic rate at
energy $E/k_B=500$~nK has been included for comparison in the leftmost panel
(dotted line).}
\label{fig4}
\end{figure}
However, the analytical result is still surprisingly
accurate for estimating the inelastic length $ \beta$, which
is a coarser quantity proportional to the overall flux reaching
the adsorbing boundary. Therefore, from~(\ref{kinel}) one
finds that ${\cal K}^{\rm inel}(q\to 0) \propto \beta$ decreases
for bosons as $a_{\rm ho}^{-2}$ whereas it increases for fermions with
$a_{\rm ho}^{2}$. As far as it concerns elastic scattering, taking the $q \to 0$
limit of~(\ref{kelb},\ref{kelf}) one finds that the ${\cal
K}_{\rm el}$ only depends on the dipolar strength $b_3$ through the
logarithmic universal term, independent of confinement. However, as in
weak confinement such zero-energy limit behavior is essentially
formal as it only holds at extremely low energies.

Examples for a finite dipole moment oriented parallel and perpendicular to
the trap axis are reported in figure~\ref{wignermd}. One can verify that
the Wigner regime is correctly reproduced by the numerical calculation
for both molecular orientations but is attained below $10^{-7}~\hbar \omega_\perp$
($10^{-4}~\hbar \omega_\perp$) for the elastic (inelastic) collision
lengths. It is also interesting to observe that at parallel polarization
$\beta$, in addition to $\alpha$, has a relatively large value of order
$a_{\rm dd}$. It is important to remark that even at parallel configuration,
in spite of the attractive character of the long-range interaction the
reaction rate does not reach the unitarity limit at low collision energy.
This is easily seen by observing that the reaction probability is simply
the reactive rate normalized to the relative velocity $v$,
and that the former quantity reduces to $4 q \beta$ for $q \to 0$. The
reaction probability therefore vanishes with collision energy due to to
the complete reflection of the incoming flux on the {\it attractive}
long-range potential tail, a purely quantum effect known as quantum
reflection. This phenomenon quantifies the deviation of
the quantum dynamics from the semiclassical limit and is known to be
supported by inverse-cubic potentials~\cite{RC-1997-PRA-1781}.

Once again, such complete quantum reflection only takes place at extremely
small collision energy, and finite energy calculations are necessary to
predict collisions rates even at the characteristic temperatures of an
ultracold gas. For instance, figure~\ref{fig4} summarizes our findings
at 5~nK. The $d=0$ reactive rate for bosons in intermediate confinement
is about 5 times larger than the corresponding weak confinement value,
in contrast with the $E \to 0$ scaling with $a_{\rm ho}^{-2}$
observed above. The fermionic $d=0$ reactive rate for intermediate
confinement about $10^2$ times larger than the corresponding weak
confinement value, in good agreement with the $a_{\rm ho}^{2}$ zero-energy law.
As in the weak confinement case, a maximum in ${\cal K}_{\rm el}^{\rm
B}$ is observed as a function of $d$, followed by a pronounced
Ramsauer-Townsend minimum where the elastic collision rate drops to
extremely small, yet nonzero, values. The Ramsauer minimum shifts at
smaller $d$ for increasing collision energy. A similar behavior is
observed for ${\cal K}_{\rm el}^{\rm F}$.

The reactive rates show a qualitatively similar behavior as in weak
confinement, comprising a maximum for bosons, and a minimum followed by a
maximum for fermions. These results qualitatively agree with the behavior
of adiabatic barrier heights.
However, due to the smaller oscillator length
here one gets more deeply in the regime $a_{\rm dd} > a_{\rm ho}$
where the reactive rates are strongly suppressed. In fact, in simple terms
a tighter transverse trap prevents more efficiently the molecules from
approaching along attractive head-to-tail reaction paths. The suppression
with $d$ becomes exponential for $a_{\rm dd} \gg a_{\rm ho}$ as implied
by semiclassical tunneling through the potential energy barrier about
the transition saddle point $\vec r^*$. 
We conclude that increasing the confinement is advantageous
for reaching a favorable rate of elastic and inelastic collision rates,
or a slow inelastic rate {\it tout court}. One may work for instance
in the minimum region of ${\cal K}^{\rm inel}$ for fermionic K-Rb, or
simply increase $d$ to large values, which is possible for more polar bosonic or fermionic
species such as Li-Cs. Unfortunately, the present results also show
that in the case of bosonic K-Rb, for which realistic induced dipole moments
are below 0.3~D, a significant suppression of ${\cal K}^{\rm inel}$
can only be attained at low temperatures of the order of few nK.

It is interesting to remark in the
leftmost panel of figure~\ref{fig2} that in the limit of large $d$ the reactive rates
for fermions and bosons tend to coincide. The numerical calculation shows
indeed not only the approximate identity of the rates but the stronger
condition on the scattering matrices $S^{\rm B} = -S^{\rm F}$.
This form of universality, often termed boson fermionization,
can be qualitatively understood as follows. The boundary condition set
on the inner sphere is universal since it completely adsorbs all flux
reaching short range irrespective of the bosonic or fermionic nature
of the colliding particles. The dynamics at sufficiently long range $
r \gg a_{\rm ho}$ is also universal because particle statistics does not
play any role in this region. Departures from universality arise in the
transition from long to the short range.
\begin{figure}[!ht]
\includegraphics[width=0.45\columnwidth,angle=-90] {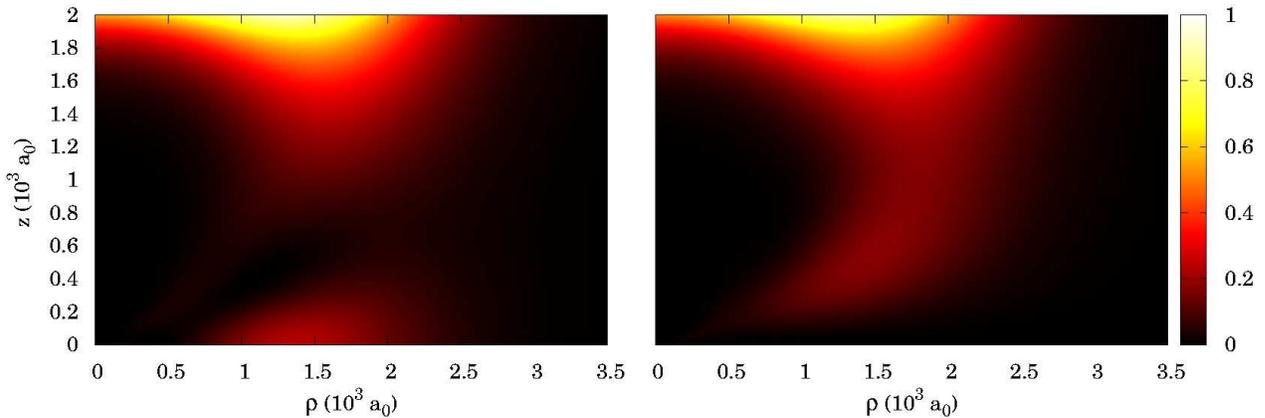}
\caption[]
{ Radial probability density $r^2 |\psi({\vec r})|^2$ (in arbitrary units) for collisions of
bosonic (right panel) and fermionic (left panel) K-Rb polar molecules in the
plane containing the trap $z$ axis and the molecular dipoles in the case of intermediate confinement. Collision
energy is low $E/k_B=5$~nK and the dipole is sufficiently strong $d=1$~D
($a_{\rm dd} \simeq 15 ~ a_{\rm ho}$) to be in the boson fermionization regime
(see text).  }
\label{fig_wf}
\end{figure}
That is, potential barrier or quantum reflection may prevent part of the
flux of the incoming wave from reaching the inner adsorbing boundary.
As the dipole increases ($a_{\rm dd} > a_{\rm ho}$), motion driven by the
dipolar interaction tends to become semiclassical and quantum reflection
is suppressed. Hence, also wave transmission inside the sphere of radius $r\approx r^*$ 
will not depend on quantum statistics. Note, however, that this does not mean that
the rate reaches the unitarity limit since the incoming wave is reflected 
by the dipolar potential at internuclear distances
$r\gtrsim r^*$. We can equivalently state that universality arises when the
transition saddle points resides in a region where molecule exchange is
irrelevant, {\it i.e.} when $r^* \gg a_{\rm ho}$. Keeping into account
the asymptotic form (\ref{psiasym}) of the wavefunction one can conclude
that fermionization $S^{\rm B} = -S^{\rm F}$ results from the combination
of the argument above {\it and} of the different sign convention for the
bosonic/fermionic asymptotic wavefunction.  Of course the elastic rates,
while related through the relation $|1-S_{jj}^{\rm B}|^2= |1+S_{jj}^{\rm
F}|^2$, do differ in general.
\begin{figure}[ht!]
\includegraphics[width=1.1\columnwidth] {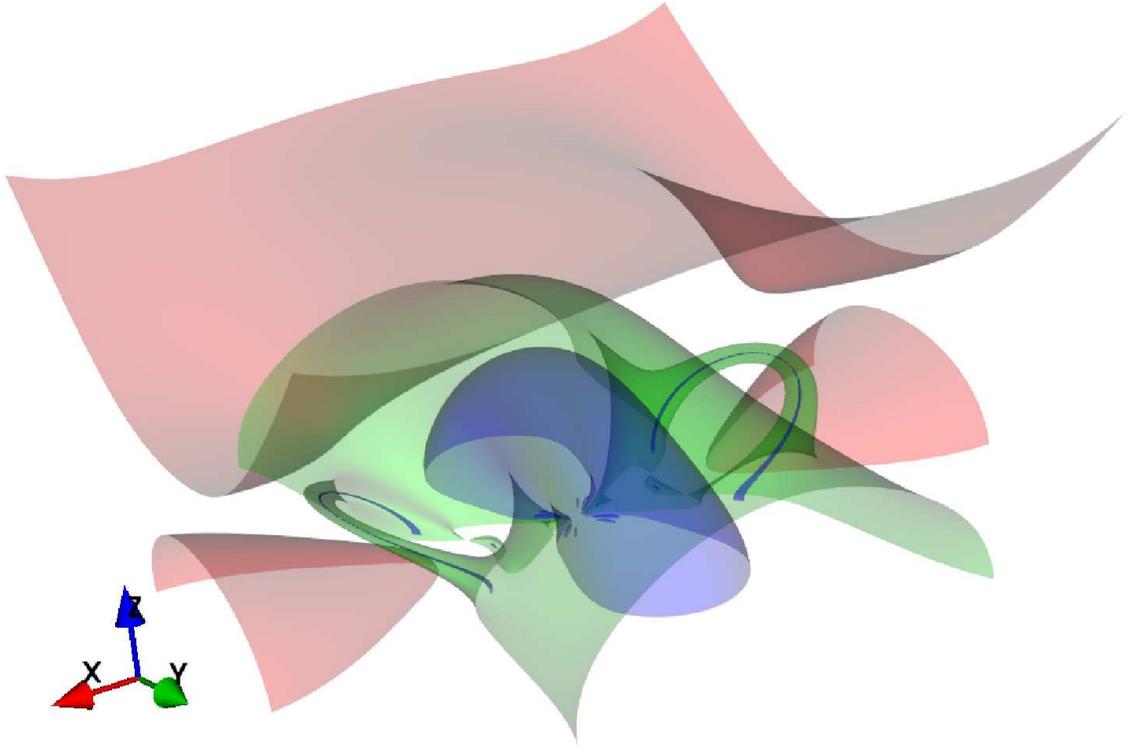}
\caption{
Isosurfaces of the probability density for the scattering bosonic wavefunction. 
The quantity $r^2 |\psi({\vec r})|^2$ is constant on each surface and varies by two order of magnitudes
from one surface to the other, increasing from colder to warmer color codes.
The wavefunction is represented in a three dimensional box of equal $1750~a_0$ sides. 
The isosurfaces are symmetric with respect to the $xz$ and $yz$ planes.
Physical parameters are as in figure~\ref{fig_wf}. 
}
\label{fig_wf_3D}
\end{figure}

It should be stressed that universality only concerns asymptotic
properties and does not imply similar dynamical behavior for bosons
and fermions at short distances. Figure~\ref{fig_wf} shows for instance
the molecule probability density in the plane of the dipoles for $r
\lesssim a_{\rm ho}$ and $d=1$~D, where one can clearly observe different patterns
for bosons and fermions with in particular the effect of particle
symmetrization making the fermionic wavefunction vanish in the $z=0$
plane.
One may also notice that the strong repulsion between parallel dipoles
pulls the molecules away from the trap axis. 

Full three dimensional analysis is represented in figure~\ref{fig_wf_3D} in the case of bosonic molecules,
where by symmetry we considered the $z>0$ half space only. 
One clearly observes that for decreasing distance the
probability distribution 
becomes quite strongly confined by
dipolar repulsion in the plane containing the dipoles and the trap axis (plane $xz$), with a vanishingly
small amplitude in the $yz$ plane.
Interestingly, one may remark the presence of a relatively large closed line (a smoke ring) where the
density probability vanishes. Inspection of the velocity field 
shows that the streamlines encircle the nodal ring, a configuration termed toroidal vortex in the seminal
paper of~\cite{1976-JOH-JCP-5456}. It is worth stressing that the presence
of the vortex is not due to a singularity in the potential but merely to quantum
interference effects.
The presence of vortices has also been observed in the context of one dimensional
quantum reactive models ~\cite{BK-1976-JOH}, although in the present case of ultracold energies the size the rings are localized
to larger distances (larger than about $3 {\bar a}$ in the case of the figure, in a region dominated by dipolar forces).
We observe the presence of a similar toroidal vortex in the fermionic case as well (not shown) with 
the notable difference that the vortex now crosses the nodal $z=0$ plane in two exceptional points. 
\begin{figure}[ht!]
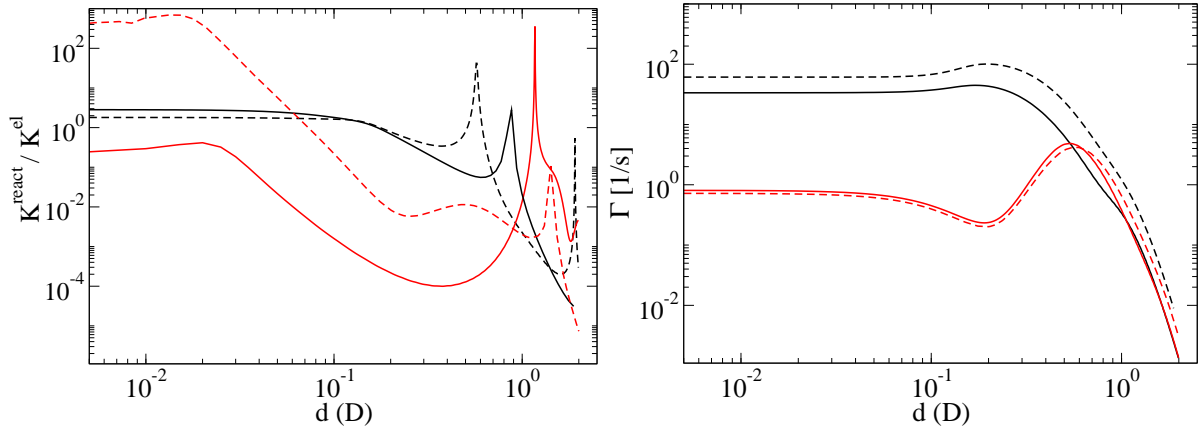

\centerline{
\epsfig{file=fig9a.eps,width=.5\columnwidth}
\epsfig{file=fig9b.eps,width=.5\columnwidth} }
\caption{
Left: the ratio between reactive and elastic collision rates for bosons (black) and fermions (red) confined in quasi-1D (full lines) or quasi-2D (dashed lines) trap with $a_{\rm ho}=10~\bar{a}$ at collision energy $E/k_B=500$nK. Right: the decay rates $\Gamma=\mathcal{K}^{\rm react} n$ for the same parameters and equivalent 3D density equal to $10^{12}$cm$^{-3}$.
}
\label{comp}
\end{figure}

To conclude, it is interesting to compare the efficiency of suppressing reactive
collisions using quasi-1D and quasi-2D confinement. We consider both
the ratio between reactive and elastic collisions
and the effective decay rate $\Gamma=\mathcal{K}^{\rm
react} n$, where $n$ is the $d$-dimensional number density measured in
${\rm cm}^{-d}$, which can be transformed to equivalent 3D density via
$n_{3D}=n/a_{ho}^{3-d}$. In a pancake-shaped trap with dipoles aligned
to repel each other, significant suppression of reaction rates has been
reported~\cite{AM-2010-PRL-073202,Quemener2011,PSJ-2011-PCCP-19114}. Here,
we find that confining the system in additional direction does not lead
to significant improvements. Figure~\ref{comp} shows indeed our results for
$a_{\rm ho}=10~\bar{a}$ at $E/k_B=500$nK for bosons and fermions. In the
case of bosons, both the ratio between reactive and elastic collision
rates and the decay rate are similar for 1D and 2D. For fermions we
find some improvement in the ratio in one-dimensional setups. However,
comparing only the rate constants does not include the impact of
many-body correlations, which induce strong suppression of reactions in
one dimension via continuous Zeno effect~\cite{Zhu2014}, especially when
a lattice potential is present along the tube direction.

\subsection{Strong confinement}
\begin{figure}[ht!]
\includegraphics[width=0.9\columnwidth] {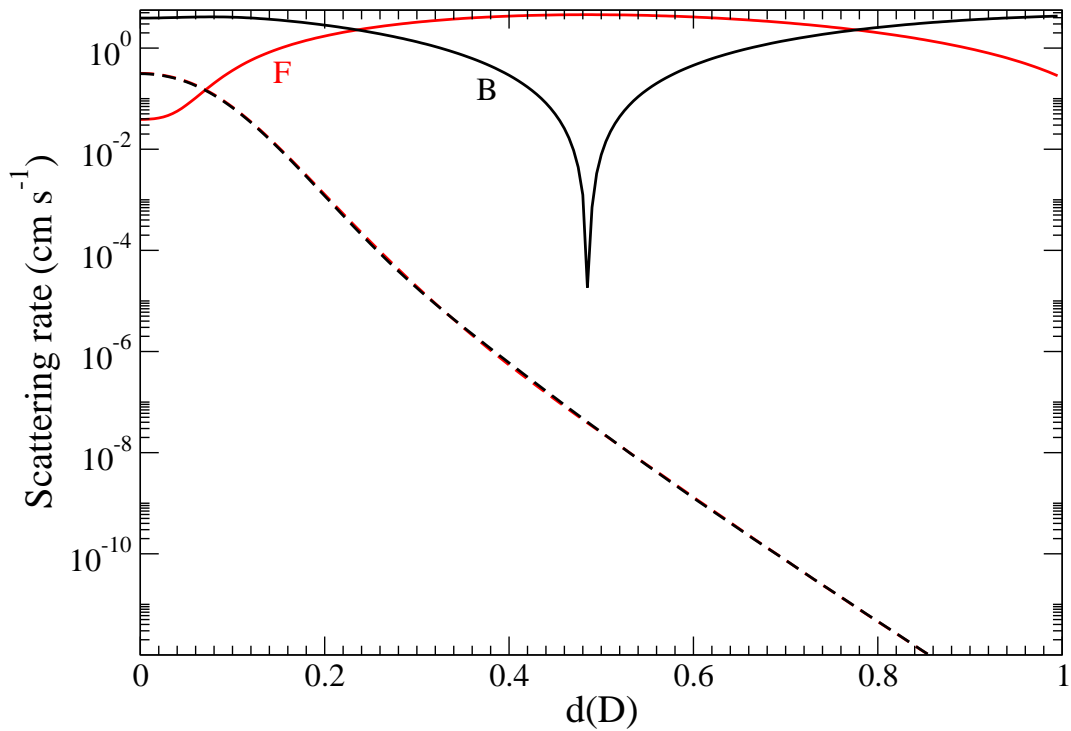}
\caption{\label{fig6}
Same as figure~\ref{fig2} but for strong confinement and $E/k_B=500$~nK.
}
\end{figure}

In the strong confinement case $a_{\rm ho}={\bar a}$ the trap is so tight
that its effect becomes important even on the short scale of the van
der Waals interaction. As shown on figure~\ref{fig6}, the elastic rates for bosons and fermions initially
differ by more than one order of magnitude for $d=0$ and become soon comparable
as a function of $d$, with the bosonic curve showing a pronounced
Ramsauer-Townsend minimum.

Quite strikingly the reactive rates for bosons and fermions are
quantitatively very similar even for $d=0$, and become exponentially
suppressed with $d$. As in the case of intermediate confinement, the
numerical calculation shows that to very good accuracy $S^{\rm B}=-S^{\rm
F}$. Moreover, wavefunction inspection (not shown) reveals that the
density probability amplitudes are very similar for bosons and fermions at
all distances, with minor differences only at $r\approx r_c$. In simple
terms we can summarize by saying that confinement is so strong that it
effectively prevents molecule exchange and under these conditions the
particle statistics becomes immaterial both with respect to the elastic
and the inelastic dynamics.  The system behaves therefore as a truly 1D
(as opposed to quasi-1D) system.

\subsection{Inelastic collisions (intermediate confinement)}
\begin{figure}[ht!]
\includegraphics[width=0.8\columnwidth] {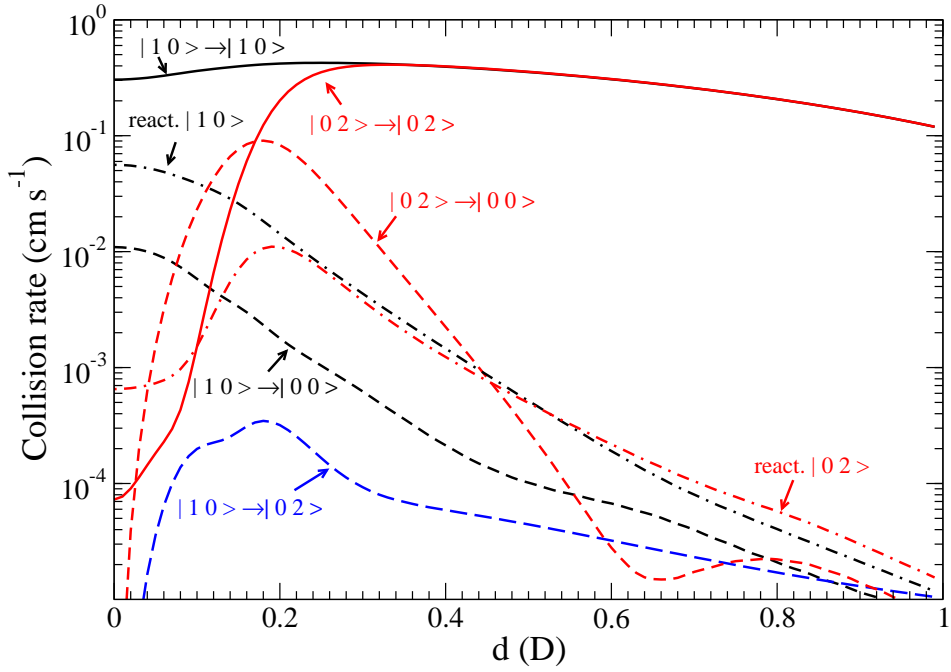}
\caption{
Elastic (full lines), superelastic (long-dashed line), inelastic (short-dashed lines), and reactive (dash-dotted lines) collision
rates for bosonic molecules in a 1D-trap as a function of the molecular dipole moment.
Different state-to-state $| n_r \Lambda \rangle \to | n_r^\prime \Lambda^\prime \rangle$ processes are separately shown for
$| 1 0 \rangle$ and $|0 2 \rangle$ initial states.
Results are for intermediate confinement and fixed collision energy
$E/k_B=5$~nK. Calculation is performed for a Hamiltonian symmetry block $\{
\pi_x,\pi_y \}=\{ 1,1 \}$ containing the ground state of the transverse oscillator.
}
\label{inelbos}
\end{figure}

If the molecules are not prepared in the ground state of the
transverse oscillator, inelastic transitions to lower trap levels
can occur even at very low collision energy. We show as an example
in figure~\ref{inelbos} the low energy collision rates for bosonic
molecules prepared in either degenerate initial states $ | 1 0 \rangle $
or $| 0 2 \rangle$.

\begin{figure}[ht!]
\includegraphics[width=0.8\columnwidth] {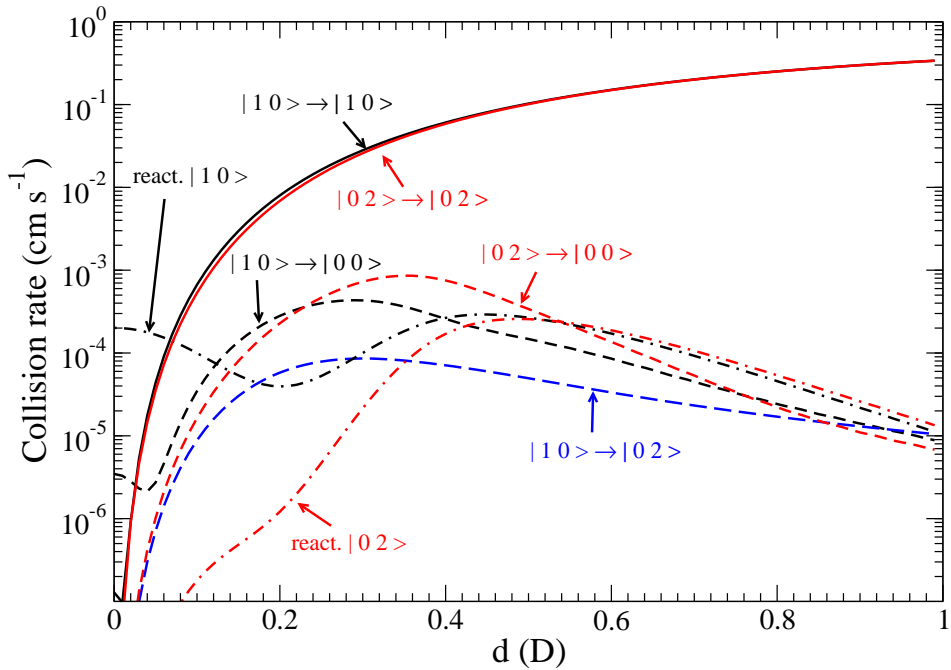}
\caption{
Same as figure~\ref{inelbos} but for fermions.
}
\label{inelfer}
\end{figure}
As a general feature, collisions starting with molecules with a relative
axial angular momentum $\Lambda >0$ experience a centrifugal barrier
in the incoming channels that pushes the molecules away from
the trap axis hence suppressing the collisional interaction. Such
effect is observed both for elastic and reactive processes, and is
very pronounced for the inelastic $|0 2 \rangle \to | 1 0 \rangle $
process. Molecule preparation in a quantum state with $\Lambda > 0$ is
therefore at least in principle a mean to control inelastic and reactive
collision rates.

It is interesting to observe in figure~\ref{inelbos} that as the dipole moment
increases above $d \approx 0.3$~D, the dipolar force dominates the centrifugal
barrier at long range, giving rise to a novel universal behavior in which
the elastic $| n_r \Lambda \rangle \to | n_r \Lambda
\rangle$ collision rates 
become independent of the initial quantum numbers. Inelastic
transitions between trap levels are found to proceed at comparable
rates as the reactive ones, both being progressively suppressed with
$d$. Finally, superelastic (energetically neutral) $ | 1 0 \rangle
\leftrightarrow | 0 2 \rangle$ collision tend to be slow at each $d$.

The fermionic case summarized in figure~\ref{inelfer} shares with the
bosonic case some qualitative features such as the drop of the inelastic
and reactive rates for large $d$. However, fermionic symmetry implies
the presence of a centrifugal barrier even for an initial $\Lambda = 0$,
such that the elastic $|1 0 \rangle \to | 1 0 \rangle$ collision
rate section is also strongly suppressed and follows in fact very closely
the evolution of the $|0 2 \rangle \to | 0 2 \rangle$ curve with $d$. The reactive and
inelastic rate are on the converse several orders of magnitudes smaller
for collisions starting in $|0 2 \rangle $ than in the $|1 0 \rangle
$ quantum state, because of the presence of a stronger effective barrier mainly
determined by the $\ell =3$ partial wave rather than $1$ as in
the case of $|1 0 \rangle $ collisions.

\subsection{Dependence on the molecular dipole orientation (intermediate confinement)}

\begin{figure}[hb!]
\includegraphics[width=0.9\columnwidth] {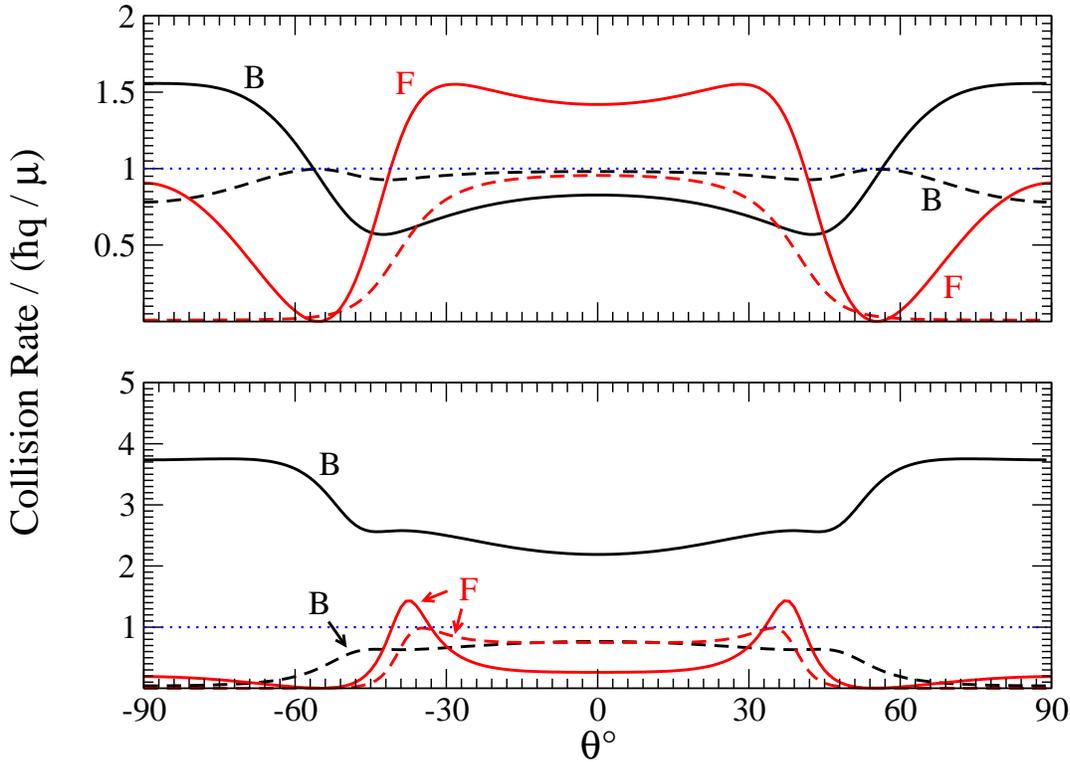}
\caption{
Collision rates as a function of the polar angle $\theta$ between $\vec d$ and
the trap axis $z$. Elastic (full lines) and inelastic (dashed lines) rates normalized
to the relative velocity are shown for bosons (B) and fermions (F) at a collision energy
of $E / k_B =5$~nK (lower panel) and 500~nK (upper panel). The dotted lines represent the upper limit of the reactive collision rates.
}
\label{fig8}
\end{figure}

Varying the orientation of the molecular electric dipoles can provide
another knob to control the collision dynamics in the experiments.
We represent in figure~\ref{fig8} the collision rates at two sample
collision energies $E/k_b=5$ and $500$~nK for bosonic and fermionic
molecules for a dipole moment such that $a_{\rm dd}=a_{\rm ho}$ (see
table~\ref{table1} for typical numerical values) in an intermediate
confinement geometry. In order to ease the comparison between the two
energies we normalize the collision rate to the relative velocity $v
$, such that the resulting reactive rate becomes the
reaction probability. As expected, the reactive collision rates reach
their minimum value when the molecular dipoles are in a repulsive
configuration orthogonal to the trap axis. The maximum takes place
at parallel configuration, where boson and fermions react at a common
rate. The lower panel of figure~\ref{fig8} demonstrates that at $5$~nK
the rate is still relatively far from the unitarity limit ${\cal K}^{\rm
react} = \hbar q / \mu$, meaning that quantum reflection is significant. Conversely, at $500$~nK (upper panel in figure~\ref{fig8}) the reactive
rate is close to unitarity.

One may also note that the elastic rate for fermions in strongly
suppressed at a specific (magic) value of the angle, which is in a
good approximation given by the condition $b_3 = 0$, corresponding to
$\theta \approx 55^\circ$. Since this value is robust against
changes in the collision energy, a non-interacting gas behavior should
be experimentally observable in a thermal sample by a suitable choice
of the direction of the polarizing electric field. 
Finally, the elastic rate for bosons
is modulated as a function of the dipole orientation but its overall
range of variation and hence the possibility of experimental control is
relatively limited at both investigated temperatures.


\section{Conclusions}
\label{conclusions}

In conclusion, we have theoretically studied polar molecule collisions
in quasi-1D geometries. Our numerical model allows us to treat dipole
strengths of experimental interest for dipoles forming an arbitrary angle with
the trap axis. As expected, polarizing the molecular electric dipoles
orthogonal to the trap axis allows one to suppress undesired reactive
rates by increasing the magnitude of $d$. Unfortunately, for moderate dipole
moments of commonly used species such as bosonic K-Rb such suppression is
only significant at very low temperatures or extremely tight confinement.
We have identified a dipolar universal regime where bosons and
fermions present equivalent asymptotic behavior at large $r$ or, in the
case of extremely tight confinement, equivalent dynamical behavior at
all distances. We also showed that in terms of suppressing the reactive collisions, 
quasi-1D confinement gives little advantage over the quasi-2D one.

In the case where the molecules populate excited trap states, inelastic
processes have been found to proceed essentially at a similar rate as the
reaction. The influence of the molecular dipole orientation on collision
is quantitatively important in particular for elastic scattering of
fermionic molecules, that become non-interacting at a magic value of
the dipole-trap axis angle.

In perspective, it may be interesting to apply the model introduced here to non-reactive
species and to extend our calculations to larger values of $d$ by relaxing the fixed-dipole
approximation.

\ack

We would like to thank G. Modugno for useful discussions.
This work was supported by the Agence Nationale de la Recherche (Contract
No. ANR-12-BS04-0020-01), Foundation for Polish Science International PhD Project co-financed by the EU European Regional Development Fund and National Center for Science grants DEC-2011/01/B/ST2/02030 and DEC-2013/09/N/ST2/0218.

\section*{References}
\bibliography{Journal,1Dmol}
\bibliographystyle{jphysicsB}

\end{document}